# Investigating Document Type, Language, Publication Year, and Author Count Discrepancies Between OpenAlex and Web of Science

Philippe Mongeon, Madelaine Hare, Poppy Riddle, Summer Wilson, Geoff Krause, Rebecca Marjoram, and Rémi Toupin

Department of Information Science, Dalhousie University

Corresponding author: Philippe Mongeon (pmongeon@dal.ca)

# 1. Abstract

Bibliometrics, whether used for research or research evaluation, relies on large multidisciplinary databases of research outputs and citation indices. The Web of Science (WoS) was the main supporting infrastructure of the field for more than 30 years until several new competitors emerged. OpenAlex, a bibliographic database launched in 2022, has distinguished itself for its openness and extensive coverage. While OpenAlex may reduce or eliminate barriers to accessing bibliometric data, one of the concerns that hinders its broader adoption for research and research evaluation is the quality of its metadata. This study aims to assess metadata quality in OpenAlex and WoS, focusing on document type, publication year, language, and number of authors. By addressing discrepancies and misattributions in metadata, this research seeks to enhance awareness of data quality issues that could impact bibliometric research and evaluation outcomes.

# 2. Introduction

Bibliometrics has been used in research and data-driven research evaluation for approximately half a century (Narin, 1976), originally supported by the development of the Web of Science[1] (WoS), which was conceptualized in 1955 and launched in 1964 (Clarivate, n.d.). It took more than 30 years for other players to emerge: Elsevier's Scopus[2] was founded in 1996, Crossref[3] in 1999, Google Scholar[4] in 2004, Microsoft Academic (now integrated into OpenAlex) and Dimensions[5] in 2018, and OpenAlex in 2022[6].

---

[1] https://www.webofscience.com/wos/
[2] https://www.scopus.com/
[3] https://www.crossref.org/

The value of bibliographic data sources is derived from different elements, including their coverage, completeness, and data accuracy (Visser et al., 2021), and may be contingent on the intended use. Major proprietary databases like Scopus and WoS are widely regarded as high-quality sources (Singh et al., 2024), but they are certainly not free of errors. Past studies have identified issues such as DOI duplication ("identity cloning") and inaccurate indexing of citations (Franceschini et al., 2015; Franceschini et al., 2016b, 2016a). Álvarez-Bornstein et al. (2017) found 12% of sampled WoS records lost funding data and showed inconsistencies across subfields. Maddi & Baudoin (2022) reported 8% of papers from 2008–2012 had incomplete author-affiliation links, improving to under 1.5% after database updates. Liu et al. (2018) noted over 20% of missing author addresses in WoS, varying by document type and year. Liu et al. (2021) outlined four causes of discrepancies between Scopus and WoS: differing publication date policies, document omissions, duplicate entries, and metadata errors. Donner (2017), for example, found document type discrepancies between WoS and journal websites or article full text in a sample of 791 publications, with letters and reviews particularly affected.

The advent of OpenAlex, an open database that indexes over 250 million scholarly works with broader coverage of the Humanities, non-English languages, and the Global South than traditional indexes (Priem et al., 2022), generated a wave of studies aimed at understanding this new source and assessing its suitability for bibliometric analyses and research evaluation. Many of these studies compared the coverage of OpenAlex to that of more established databases. Culbert et al. (2024) investigated the coverage of reference items between OpenAlex, WoS, and Scopus. They found that OpenAlex was comparable with commercial databases from an internal reference coverage perspective if restricted to a core corpus of publications similar to the other two sources, though it lacked cited references. Alonso-Alvarez & Eck (2024) also found missing references. OpenAlex was found to lack funding metadata (Schares, 2024) and institutional affiliations (Bordignon, 2024; Zhang et al., 2024) but to have higher author counts compared to WoS and Scopus (Alonso-Alvarez & Eck, 2024). Haupka et al. (2024) observed that a broader range of materials were classified as articles in OpenAlex compared to Scopus, WoS, and PubMed, potentially explained by OpenAlex's reliance on Crossref's less granular system of classification (Ortega & Delgado-Quirós, 2024). At the journal level, Simard et al. (2025) and Maddi et al. (2024) found that OpenAlex provides a higher and more geographically balanced coverage of open access journals compared to WoS and Scopus. Céspedes et al. (2024) also found that OpenAlex had a broader linguistic coverage (75% English) than WoS (95% English). They also identified discrepancies in OpenAlex's language metadata, noting that approximately 7% of articles were incorrectly classified as English, mainly due to the algorithmic detection of language from the title and abstract metadata in OpenAlex.

OpenAlex's openness and broad coverage of the scholarly record have likely contributed to its fast adoption as a data source for scientometric research. There is evidence that it is indeed suitable in certain cases, such as large-scale, country-level analyses (Alperin et al., 2024).

---

[4] https://scholar.google.com/
[5] https://www.dimensions.ai/
[6] https://openalex.org/

However, there is some warranted resistance to its adoption for formal research evaluation purposes, given its known imperfections, such as partial correctness or erroneous metadata (Haunschild & Bornmann, 2024). The Centre for Science and Technology Studies (CWTS) was, to our knowledge, the first to make such a use of OpenAlex. Since 2023, they have been using it to produce the open edition[7] of their now well-established CWTS Leiden Ranking.

The CWTS Leiden Ranking Open Edition employed several strategies to mitigate the limitations of the OpenAlex database and the risks associated with its use. It was limited to approximately 9.3 million articles and reviews that are in English, are published in a set of core, international journals and cite or are cited in other core journals (see Eck et al., 2024 and the ranking website for more details). The objective was to reproduce the WoS-based Leiden Ranking as closely as possible. As such, the open ranking does not take full advantage of the more extensive coverage of OpenAlex. In their blog post, Eck et al. (2024) provide some insights into the publications included in the two versions of the ranking: 2.5 million publications are included only in the Open Edition. In most cases (1.7 million publications) this is because they are not covered in WoS. On the other hand, 0.7 million publications are only included in the WoS-based ranking, explained by publication year discrepancies between the two databases, leading to some articles being included in one ranking but not the other, and by the missing institutional affiliations in OpenAlex records (Eck et al., 2024). The challenges of determining the publication year of an article were also previously discussed by Haustein et al. (2015).

Because of the large overlap in the data for the two rankings, they produced similar outcomes, with the same institutions ranking on top in the open edition. Yet, there were some differences with some universities moving up or down by a few ranks. For example, Dalhousie University ranked #15 for Canadian universities in the WoS version of the ranking, and #13 in the OpenAlex version. Being presented with two rankings that used the same process but provide different outcomes due to their data sources may lead to the dilemma of Segal's law, according to which "A man with a watch knows what time it is. A man with two watches is never sure" (Liu et al., 2021).

## 2.1 Research objectives

Our study was partly inspired by the launch of the CWTS Leiden Ranking Open Edition and its differences with the WoS-based edition. However, our intent is not to compare the two rankings and explain the differences in their outcome, but to compare metadata records for the subset of publications that are indexed in both OpenAlex and WoS, with a focus on describing discrepancies and measuring their frequency. We do not consider WoS as a gold standard against which OpenAlex records should be tested for their quality. Instead, we consider each database as a benchmark for the other. Our research questions are as follows:

- **RQ1**: How frequent are discrepancies between WoS and OpenAlex records in terms of document type, language, publication year, and number of authors?
- **RQ2**: What share of records with discrepancies are correct in WoS, OpenAlex, neither, or both?
- **RQ3**: What explains these discrepancies?

---

[7] https://open.leidenranking.com/

The choice of the metadata elements selected for the study (document type, language, publication year, and number of authors) was guided by their use as factors for inclusion in the CWTS Leiden Ranking: articles and reviews (document type) in English (language) published in a three-year period (publication year). The same metadata are also relevant in the production of normalized indicators for bibliometric research more generally. Indeed, normalized indicators would typically aim to compare publications of the same type, published in the same year. Additionally, the number of authors tends to be the denominator when counting publications using the fractional counting method (also provided in the Leiden Ranking). Therefore, discrepancies in these metadata elements of publication records are more likely to have a direct effect on the outcome of bibliometric analysis.

To our knowledge, our study is the first to compare the metadata quality of the document type, language, publication year, and author count of the subset of publications indexed in both OpenAlex and WoS. Our findings may offer useful insights to those using or planning to use OpenAlex. They may also point to possible paths to correct metadata errors, which could be helpful for database providers. Our intent is not to establish best practices formulate recommendations for database creators or users, nor to determine whether one database is better than another. Our study is solely motivated by the desire to contribute to a better understanding of our data sources.

## 3. Data and methods

### 3.1 Data collection

The WoS data used in this study was retrieved from a relational database version of WoS hosted by the Observatoire des sciences et des technologies (OST) and limited to the Science Citation Index (SCI), the Social Sciences Citation Index (SSCI), and the Arts & Humanities Citation Index (A&HCI). We collected all WoS records with a DOI published between 2021 and 2023 (N = 7,661,474). We removed 30,3094 (0.4%) WoS records with multiple document types to avoid complications with the analysis. Of the remaining 7,631,080 WoS records, 6,599,479 (86.5%) had a DOI match in the February 2024 snapshot of OpenAlex accessed through Google Big Query (see Mazoni & Costas, 2024). Conversely, we collected all OpenAlex records with a DOI and a publication year between 2021 and 2023 (N = 26,942,411). We found 13,618 additional matches to WoS records published outside of the 2021-2023 period, for a total of 6,613,097 records matched. We used the subset of 6,608,243 Web of Science records with a single DOI match in OpenAlex for our analysis. In the OST database, every journal is assigned to one of 143 specialties and 14 disciplines of the National Science Foundation (NSF) classification.

### 3.2 Identification of discrepancies

For each matching WoS and OpenAlex record, we compared the following four metadata elements: 1) document type, 2) language, 3) publication year, and 4) number of authors.

For the document type, we did not consider discrepancies where a record is a review according to WoS and an article according to OpenAlex, and vice versa (i.e., we consider articles and reviews as the same document type). Furthermore, OpenAlex indexes conference papers as articles, and the source type (conference) is meant to distinguish them from journal articles. For these reasons, we only considered discrepancies for which the record is identified as an article or a review in WoS but not in OpenAlex, or in OpenAlex but not in WoS. We also excluded discrepancies for which the record is a meeting abstract in WoS and an article in OpenAlex, because OpenAlex classifies conference proceedings as articles and relies on the source information to distinguish the two. Overall, we found 429,692 discrepancies that met these criteria (6.5% of all records in the dataset).

The identification of discrepancies in language, publication year, and number of authors was limited to the subset of 5,936,235 articles and reviews with no discrepancy in document type (i.e., the records that are articles or reviews in both WoS and OpenAlex). The publication language is recorded in its long form (e.g., English) in our WoS data and its ISO code (e.g., en) in OpenAlex, so we created a list of all language combinations (e.g., English-EN, English-FR), determined which ones constitute discrepancies, and identified 33,516 discrepancies (0.6% of all articles and reviews) in publication language between WoS and OpenAlex.

The publication year and number of authors being numeric indicators, identifying discrepancies in these metadata fields is easily done by calculating the difference between the two values and flagging all non-zero values. We identified a total of 480,884 discrepancies (8.1% of all articles and reviews) in publication years and 71,133 discrepancies (1.2% of all articles and reviews) in the number of authors. We recorded them both as dichotomous variables (discrepancy vs no discrepancy) and as numeric differences between the two values, which allowed us to measure both the frequency and the strength of the discrepancies.

We obtained the direction and extent of the discrepancies by substracting the OpenAlex publication year or number of author from the WoS values. A difference of 1 indicates that according to WoS the article was published one year (or has one more author) later than according to OpenAlex. Conversely, a difference of -1 indicates that the article was published one year earlier (or has one fewer author) according to WoS than according to OpenAlex. We then grouped cases in the bins (-4 or less, -2 to -3, -1, 1, 2 to 3, 4 or more).

## 3.3 Investigation of discrepancies

We took a random sample of discrepancies for manual investigation. For each type of discrepancy, we calculated the required sample size with a 95% confidence level and a margin of error of ±5%. The sample discrepancies were manually investigated by looking at the article on the journal's website and the full text when necessary and available. We recorded whether the WoS or OpenAlex record was correct and, when possible, explained the discrepancy. For the publication year, it is typical for articles to have two publication dates: the date of the first online publication and the date of the publication of the issue. For cases in which the online publication year was recorded in one database and the issue year was recorded in the other, we did not consider either database to be correct.

For author discrepancies, the landing page and, when available, the published version were examined for author counts. Where available, author declarations, acknowledgements, and CRediT declarations were also considered to clarify authorship where groups or consortia were involved in the production of the work. For languages, the full text was the primary source, followed by the publisher's landing page for recording observations on language assignment. We recorded multiple language full texts if available, as well as translated abstracts available on the landing page. There are limitations to using the publisher landing page with a web browser. The language encoding header of the browser's HTTP GET request could be used by the web server to display the user's preferred language type, if the server is configured to respond to such a request.

This process allowed us to go beyond the simple counting of differences between the databases and gain insight into the different factors that can cause discrepancies and estimate the percentage of erroneous records in each database. One caveat, of course, is that because our process relies on discrepancies to identify errors, we are not considering cases where both databases contain the same error.

# 4. Results

We structured our results section by type of discrepancy. For each, we first show the distribution of values in the Web of Science and the OpenAlex datasets. Note that these are not the distributions in the entire Web of Science OpenAlex databases, but the distributions in the subset of records with a DOI match between databases. Second, we present descriptive statistics related to the discrepancies identified and the sample that was used for investigations. Finally, we present the results of these investigations.

## 4.1 Document type

### 4.1.1 Document type distribution in WoS and OpenAlex

Table 1 and Table 2 present the distribution of records across document types in WoS and OpenAlex, respectively, to provide a general picture of the databases' content and the differences in their classification. While WoS contains twice as many document types as OpenAlex, these differences appear mainly among the less frequent types, in line with the findings of Haupka et al. (2024). Most documents in both data sources are articles and reviews.

*Table 1: Distribution of records by document type in WoS*

| WoS Document Type | n | Records pct |
|---|---|---|
| article | 5,434,906 | 82.24 |
| review | 510,322 | 7.72 |
| editorial material | 266,705 | 4.04 |
| meeting abstract | 117,136 | 1.77 |
| letter | 113,874 | 1.72 |
| book review | 67,571 | 1.02 |
| correction | 67,271 | 1.02 |
| news item | 11,983 | 0.18 |
| retraction | 8,290 | 0.13 |
| biographical-item | 6,516 | 0.10 |
| Other* | 3,669 | 0.06 |
| total | 6,608,243 | 100.00 |

*Includes cc meeting heading, poetry, expression of concern, reprint, art exhibit review, item withdrawal, film review, bibliography, fiction, creative prose, theater review, record review, music performance review, software review, music score review, hardware review, tv review, radio review, dance performance review, excerpt, database review, data paper, note, and script. It should be noted that TV Review, Radio Review, Video Review were retired as document types and are no longer added to items indexed in the WoS Core Collection. They are still usable for searching or refining/analyzing search results.

*Table 2: Distribution of records by document type in OpenAlex*

| | Records | |
|---|---:|---:|
| **OpenAlex Document Type** | **n** | **pct** |
| article | 5,845,248 | 88.45 |
| review | 511,686 | 7.74 |
| letter | 136,051 | 2.06 |
| editorial | 59,662 | 0.90 |
| erratum | 48,189 | 0.73 |
| retraction | 4,704 | 0.07 |
| book-chapter | 1,333 | 0.02 |
| preprint | 872 | 0.01 |
| paratext | 307 | 0.00 |
| other | 96 | 0.00 |
| Other* | 95 | 0.00 |
| total | 6,608,243 | 100.00 |

*Includes report, dataset, other, dissertation, supplementary-materials, and reference-entry.

Table 3 and Table 4 show that the vast majority, 303,582 (97.1%), of the 312,580 discrepancies are cases where a record is an article or review in OpenAlex but not WoS. We find that, in both tables, the majority of discrepancies indicate a misclassification of a record as an article or review. However, WoS is much more accurate in detecting misclassifications in OpenAlex (93.5% of the sample) than OpenAlex is in WoS (68.8% of the sample). This generally points to a much larger number of OpenAlex records misclassified as articles or reviews compared to WoS.

*Table 3: Discrepancies for documents categorized as articles or reviews in the WoS but not in OpenAlex*

| Type in OpenAlex | Discrepancies | | Sample | | Errors | |
|---|---|---|---|---|---|---|
| | n | pct | n | pct | n | pct |
| letter | 5,008 | 55.7 | 331 | 56.7 | 219 | 66.2 |
| editorial | 1,341 | 14.9 | 87 | 14.9 | 63 | 72.4 |
| book-chapter | 1,304 | 14.5 | 85 | 14.6 | 85 | 100.0 |
| preprint | 840 | 9.3 | 54 | 9.2 | 14 | 25.9 |
| paratext | 240 | 2.7 | 14 | 2.4 | 14 | 100.0 |
| erratum | 149 | 1.7 | 8 | 1.4 | 6 | 75.0 |
| other | 83 | 0.9 | 4 | 0.7 | 0 | 0.0 |
| retraction | 28 | 0.3 | 1 | 0.2 | 1 | 100.0 |
| book | 5 | 0.1 | 0 | 0.0 | 0 | 0.0 |
| total | 8,998 | 100.0 | 584 | 100.0 | 402 | 68.8 |

*Table 4: Discrepancies for documents categorized as articles or reviews in OpenAlex but not in WoS*

| Type in WoS | Discrepancies | | Sample | | Errors | |
|---|---|---|---|---|---|---|
| | n | pct | n | pct | n | pct |
| editorial material | 161,628 | 53.2 | 350 | 54.3 | 308 | 88.0 |
| book review | 67,435 | 22.2 | 146 | 22.6 | 146 | 100.0 |
| letter | 30,699 | 10.1 | 65 | 10.1 | 65 | 100.0 |
| correction | 19,303 | 6.4 | 40 | 6.2 | 40 | 100.0 |
| news item | 11,336 | 3.7 | 23 | 3.6 | 23 | 100.0 |
| biographical-item | 5,941 | 2.0 | 11 | 1.7 | 11 | 100.0 |
| retraction | 3,727 | 1.2 | 7 | 1.1 | 7 | 100.0 |
| other | 3,513 | 1.2 | 3 | 0.5 | 3 | 100.0 |
| total | 303,582 | 100.0 | 645 | 100.0 | 603 | 93.5 |

## 4.2 Publication Language

### 4.2.1 Language Distribution in WoS and OpenAlex

Table 5 and Table 6 show the distribution of articles and reviews across languages in WoS and OpenAlex, respectively, limited to the top 10 most frequent languages. Unsurprisingly, the quasi totality of records is in English in both databases. It is important to note, once again, that our results are not meant to reflect the composition of the entire databases, but only the subset of documents included in both. This is expected to underestimate the representation of non-English literature, especially on the OpenAlex side (Simard et al., 2024). Also important to reiterate here is that our WoS data includes only the SCI-E, SSCI, and A&HCI, and not the ESCI, which would include more non-English records.

*Table 5: Distribution of records by language in WoS*

| | | Records |
|---|---|---|
| **Language in WoS** | n | pct |
| English | 5,895,515 | 99.31 |
| German | 14,453 | 0.24 |
| Spanish | 7,969 | 0.13 |
| French | 6,172 | 0.10 |
| Chinese | 3,915 | 0.07 |
| Russian | 1,978 | 0.03 |
| Portuguese | 1,526 | 0.03 |
| Polish | 873 | 0.01 |
| Japanese | 773 | 0.01 |
| Turkish | 580 | 0.01 |
| Other | 2,481 | 0.04 |
| Total | 5,936,235 | 100.00 |

Aside from the dominance of English in the dataset, we also observe differences in the languages that compose the top 10 in each database, as well as in their ranking. Spanish, French, and Portuguese are much more prevalent in the OpenAlex records than in the WoS. On the other hand, Chinese is the 5th most prevalent language in the WoS data, but it does not make the top 10 in OpenAlex, in which we found only 394 articles labelled as Chinese. Since these are supposed to be the same set of records, these differences point to significant discrepancies in the language field between the two databases.

*Table 6: Distribution of records by language in OpenAlex*

| | | Records |
|---|---|---|
| **Language in OpenAlex** | n | pct |
| English | 5,886,864 | 99.17 |
| Spanish | 13,777 | 0.23 |
| German | 13,244 | 0.22 |
| French | 10,466 | 0.18 |
| Portuguese | 3,203 | 0.05 |
| Polish | 1,058 | 0.02 |
| Turkish | 909 | 0.02 |
| Hungarian | 841 | 0.01 |
| Croatian | 591 | 0.01 |
| Romanian | 537 | 0.01 |
| Other | 3,653 | 0.06 |
| Unknown | 1,092 | 0.02 |
| Total | 5,936,235 | 100.00 |

### 4.2.2 Language Discrepancies

For this part of the analysis, we focus on articles and reviews that are in English in WoS or in OpenAlex, and in a language other than English in the other database. We do not investigate discrepancies between Portuguese and Spanish or French and German, but only discrepancies for which one of the languages is English (e.g., English-German, English-Chinese).

Table 7 and Table 8, respectively, display the number of discrepancies found for articles labelled as English in WoS and articles labelled as English in OpenAlex. They also display the number of records manually investigated to determine whether the discrepancies stem from mislabeling non-English content as English, as well as the result of that investigation. To understand how an incorrect assignment may have occurred, we looked at the abstracts and full text on the landing page as well as the full text when it was available. Most sampled works seem to have multi-language abstracts either on the landing page or in the PDF of the full text. Also, some articles are published in multiple languages, meaning that for some of the discrepancies observed, both OpenAlex and WoS are correct.

A first observation is that WoS appears to be more accurate than OpenAlex in terms of publication language. Table 7 shows that only 20,211 English publications in WoS were in another language according to OpenAlex, and in slightly less than half (49.6%) of the verified cases, the article was in fact in English.

*Table 7: Language discrepancies for articles and reviews in English in WoS*

| Language in OpenAlex | Discrepancies | | Sample | | Errors | |
|---|---:|---:|---:|---:|---:|---:|
| | n | pct | n | pct | n | pct |
| Spanish | 7,029 | 34.8 | 128 | 38.0 | 82 | 64.1 |
| French | 5,580 | 27.6 | 101 | 30.0 | 31 | 30.7 |
| Portuguese | 2,111 | 10.4 | 38 | 11.3 | 23 | 60.5 |
| German | 1,598 | 7.9 | 26 | 7.7 | 8 | 30.8 |
| Romanian | 529 | 2.6 | 8 | 2.4 | 0 | 0.0 |
| Hungarian | 520 | 2.6 | 9 | 2.7 | 9 | 100.0 |
| Turkish | 464 | 2.3 | 6 | 1.8 | 5 | 83.3 |
| Croatian | 400 | 2.0 | 4 | 1.2 | 0 | 0.0 |
| Polish | 290 | 1.4 | 3 | 0.9 | 3 | 100.0 |
| Other | 1,690 | 8.4 | 14 | 4.2 | 6 | 42.9 |
| Total | 20,211 | 100.0 | 337 | 100.0 | 167 | 49.6 |

However, we notice that for some languages, such as French, German, Romanian, and Croatian, the discrepancies correctly flagged articles mislabeled as English in WoS, although the sample sizes for Romanian and Croatian were quite small. On the other hand, the majority of articles labelled as English in WoS and as Spanish, Portuguese, or Turkish in OpenAlex were actually found to be in English.

*Table 8: Language discrepancies for articles and reviews in English in OpenAlex*

| Language in WoS | Discrepancies | | Sample | | Errors | |
|---|---|---|---|---|---|---|
| | n | pct | n | pct | n | pct |
| Chinese | 3,307 | 26.6 | 105 | 31.8 | 102 | 97.1 |
| German | 2,589 | 20.8 | 45 | 13.6 | 37 | 82.2 |
| Russian | 1,723 | 13.8 | 49 | 14.8 | 46 | 93.9 |
| French | 1,336 | 10.7 | 32 | 9.7 | 32 | 100.0 |
| Spanish | 1,200 | 9.6 | 20 | 6.1 | 18 | 90.0 |
| Japanese | 620 | 5.0 | 23 | 7.0 | 23 | 100.0 |
| Portuguese | 411 | 3.3 | 13 | 3.9 | 10 | 76.9 |
| Turkish | 131 | 1.1 | 6 | 1.8 | 6 | 100.0 |
| Polish | 92 | 0.7 | 3 | 0.9 | 3 | 100.0 |
| Other | 1,036 | 8.3 | 34 | 10.3 | 33 | 97.1 |
| Total | 12,445 | 100.0 | 330 | 100.0 | 310 | 93.9 |

## 4.3 Publication year

### 4.3.1 Publication Year Distributions in WoS and OpenAlex

Now turning our attention to discrepancies in publication years for articles in WoS and OpenAlex, we first present a statistical summary of the publication year in OpenAlex and WoS (Table 9).

*Table 9: Descriptive statistics summary of the publication year for articles and reviews in OpenAlex and WoS*

| Database | n | mean | sd | var | q1 | median | q3 | min | max | skew | kurtosis |
|---|---|---|---|---|---|---|---|---|---|---|---|
| OpenAlex | 5,936,235 | 2,021.9 | 0.952 | 0.905 | 2,021 | 2,022 | 2,023 | 1,915 | 2,025 | -7.433 | 537.6 |
| WoS | 5,936,235 | 2,022.0 | 0.820 | 0.673 | 2,021 | 2,022 | 2,023 | 1,998 | 2,023 | -0.055 | -0.5 |

### 4.3.2 Publication Year Discrepancies

We observe a high prevalence of publication year discrepancies for articles and reviews between WoS and OpenAlex, with 470,107 cases identified (see Table 10). We obtain the direction and extent of the discrepancies by subtracting the OpenAlex publication year from the WoS publication year. A difference of 1 indicates that the article was published one year later in WoS than according to OpenAlex. Conversely, a difference of -1 indicates that the article was published one year earlier according to WoS than according to OpenAlex.

*Table 10: Number of publication year discrepancies for articles and reviews in WoS and OpenAlex (total and sampled)*

| Difference in publication year (WoS - OpenAlex) | Discrepancies | | Sample | |
|---|---|---|---|---|
| | n | pct | n | pct |
| -2 to -3 | 2,549 | 0.5 | 20 | 5.1 |
| -1 | 58,526 | 12.2 | 38 | 9.6 |
| 1 | 392,681 | 81.7 | 276 | 70.1 |
| 2 to 3 | 23,166 | 4.8 | 20 | 5.1 |
| < -3 | 2,069 | 0.4 | 20 | 5.1 |
| Total | 480,884 | 100.0 | 394 | 100.0 |

### 4.3.3 Explanation for Publication Year Discrepancies

From our investigation of the discrepancies, we observed four general possible publication years that one record can have in either database:

- Issue year: the year of the journal issue in which the article is officially published. This is also the year included in the reference to the article.
- First publication year: the year in which the article was first published online before it was assigned to an issue.
- Other year: another year that is recorded in the article's history but is not a publication year (e.g., submission year, accepted year).
- Source unclear: we use this code when the publication year in the WoS or OpenAlex record does not match any of the years listed in the landing page or the PDF.

We used these four scenarios to group publication year discrepancies between WoS and OpenAlex into six categories. The first two are 1) OA has first year, WoS has issue year, and 2) OA has issue year, WoS has first year. In these cases, we considered both databases to be correct because both dates are actually referring to a publication event. The following four categories indicate errors in one of the databases that include none of the two accepted publication years: 3) OA has other year, 4) OA source unclear, 5) WoS has other year, and 6) WoS source unclear.

As shown in Figure 1, erroneous publication years are rare overall in the sample that we analyzed and mostly occur when there are significant discrepancies between the WoS and OpenAlex publication years. However, these errors appear to be more prevalent in OpenAlex. For half of the sampled articles published 2 or 3 years later according to OpenAlex than according to WoS, the source of the OpenAlex publication year was not found on the article landing page or PDF and appeared to be wrong. This was also the case for 20% of the articles with a publication year in OpenAlex three or more years earlier than in WoS. In contrast, we only found a few instances (four in total) where the WoS publication year appeared to be erroneous.

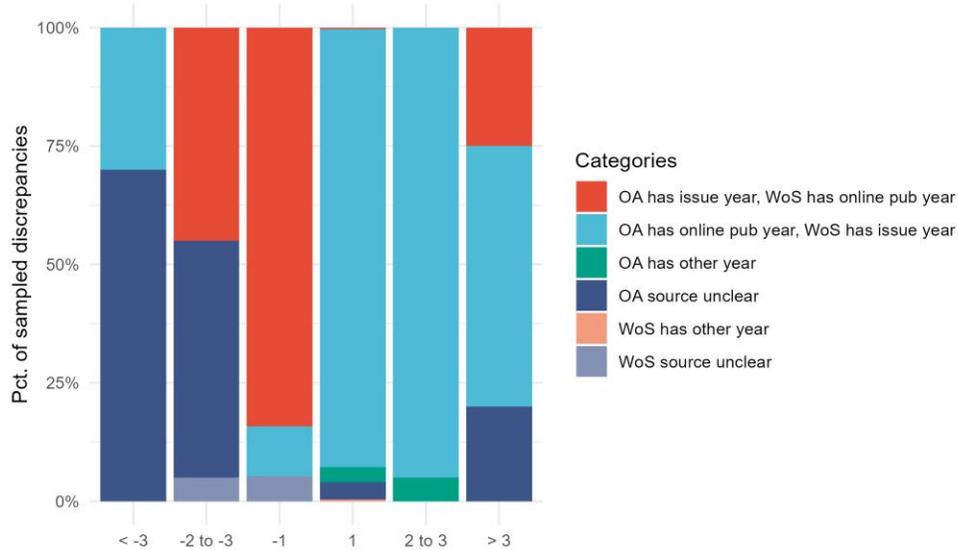

Figure 1. Relative distribution of discrepancies categories across bins of publication year differences

Figure 1 shows that discrepancies for which the OpenAlex year is prior to the WoS year are largely explained by cases where OpenAlex has the first publication year and WoS has the issue year, especially when the discrepancy is small (one year). This is expected given that the first publication year is typically before the journal issue year. Inversely, most cases in which the WoS publication year is prior to the OpenAlex publication year are explained by cases where WoS has the first publication year and OpenAlex has the journal issue year. Interestingly, neither database appears to be systematic in the way it indexes the year of first publication or the year of the journal issue.

## 4.4 Number of authors

Now turning our attention to discrepancies in publication years for articles in WoS and OpenAlex, we first present a statistical summary of the number of authors for articles and reviews in OpenAlex and WoS (Table 11) and of the number of discrepancies identified (Table 12).

Table 11: Descriptive statistics summary of the number of authors for articles and reviews in OpenAlex and WoS

| Database | n | mean | sd | var | q1 | median | q3 | min | max | skew | kurtosis |
|---|---|---|---|---|---|---|---|---|---|---|---|
| OpenAlex | 5,924,459 | 6.09 | 16.249 | 264.022 | 3 | 5 | 7 | 0 | 4,548 | 124.997 | 20,090.83 |
| WoS | 5,924,459 | 6.00 | 14.835 | 220.074 | 3 | 5 | 7 | 1 | 5,552 | 154.146 | 29,546.77 |

Table 12: Number of discrepancies in the number of authors for articles and reviews in WoS and OpenAlex (total and sampled)

| Difference in number of authors (WoS - OpenAlex) | Discrepancies | | Sample | |
|---|---:|---:|---:|---:|
| | n | pct | n | pct |
| -2 to -3 | 1,717 | 2.4 | 8 | 3.1 |
| -1 | 9,446 | 13.3 | 49 | 18.7 |
| 1 | 38,660 | 54.3 | 97 | 37.0 |
| 2 to 3 | 7,377 | 10.4 | 37 | 14.1 |
| < -3 | 9,045 | 12.7 | 47 | 17.9 |
| Total | 71,133 | 100.0 | 262 | 100.0 |

4.4.1 Explanation for Author Count Discrepancies

Of the 257 publications with discrepancies investigated, about half (122; 47.5%) had a group author on the byline. For these cases, many of the discrepancies are explained by what appears to be a different treatment of group authors by OpenAlex and WoS. As shown in Table 13, OpenAlex appears to be counting as authors each member of the group listed on the byline, but only when that list is displayed on the landing page. However, the group tends to be ignored when its members are not listed on the landing page. WoS usually does not count members of a group listed on the byline as individual authors, regardless of whether or not the names appear on the landing page.

*Table 13: Treatment of group authors in WoS and OpenAlex relative to the presence of the list of group members on the landing page.*

| Group authors on landing page | OpenAlex | | WoS | |
|---|---:|---:|---:|---:|
| | Counted | Ignored | Counted | Ignored |
| No | 4 | 65 | 1 | 68 |
| Yes | 37 | 10 | 8 | 39 |

While group members are typically not counted as authors in WoS, the group itself typically is. For example, a paper with two authors and a group with 10 members will have an author count of 3 in WoS. On the other hand, OpenAlex does not include the group in the author count, so the same paper will typically have an author count of 2 or 12, depending on whether the group members appeared on the landing page.

For papers with a group author, OpenAlex and WoS having a different author count for the same paper does not mean that either database contains the correct number of authors. However, for publications that do not have group authors, determining which database has the correct author count is more straightforward. As shown in Table 14, OpenAlex was more likely to be correct (74.8% of the time) in these cases.

*Table 14: Number and share of valid records in OpenAlex and WoS in cases of author count discrepancies with no group authors listed on the byline.*

| bin | Sample n | pct | OpenAlex n | pct | WoS n | pct |
| --- | --- | --- | --- | --- | --- | --- |
| < -3 | 6 | 4.7 | 5 | 83.3 | 1 | 16.7 |
| -2 to -3 | 8 | 6.3 | 7 | 87.5 | 1 | 12.5 |
| -1 | 44 | 34.6 | 39 | 88.6 | 5 | 11.4 |
| 1 | 26 | 20.5 | 23 | 88.5 | 3 | 11.5 |
| 2 to 3 | 27 | 21.3 | 19 | 70.4 | 8 | 29.6 |
| > 3 | 16 | 12.6 | 2 | 12.5 | 14 | 87.5 |
| Total | 127 | 100.0 | 95 | 74.8 | 32 | 25.2 |

# 5. Discussion and conclusion

## 5.1 Summary of findings

In some regards, our findings generally offer support to past claims that metadata quality in OpenAlex has room for improvement. It is particularly the case for document types, with more than 300,000 cases where a publication was classified as an article or review in OpenAlex but not in WoS, and almost all manually verified instances pointed to a true misclassification in OpenAlex. This issue may be inherited from OpenAlex's use of the document recorded in Crossref[8], but remain a potential cause for concern as document type classifications are critical for bibliometric research and evaluation. Document type misclassifications were also present in WoS, although they were much less frequent.

Discrepancies in publication language were also observed but were less common by an order of magnitude. In this case, we also did not identify a clear advantage for one database over the other. There were more cases of articles labelled as English in WoS and not in OpenAlex than the other way around; however, WoS was actually correct 50.4% of the time. On the other hand, OpenAlex had fewer articles in English that were in another language according to WoS, but those cases tended to be true misclassifications in OpenAlex. In several instances, both databases were technically correct because we found that the article was, in fact, published in multiple languages. This highlights a potential issue with metadata schemata that only allow a single value for language. The language discrepancies observed may also be explained by the fact that the OpenAlex language field is algorithmically derived from the metadata (title & abstract). The lack of standardized schemes for the romanization of non-Latin scripts may also cause some issues, not only in terms of metadata quality and the ability for algorithms to properly detect language (Shi et al., 2025).

Our investigation of publication year discrepancies highlighted that the year of first publication tended to be preferred by OpenAlex, while the use of the issue year was more frequent in WoS. However, both databases displayed inconsistencies in that regard. This challenge with the ambiguity of the publication year and the different, arguably equally valid, approaches to recording it has been previously identified and discussed by Haustein et al. (2015) and Liu et al. (2021). This is another limitation that could potentially be fixed by recording all the years in the metadata schema rather than allowing only one publication year.

---

[8] See the OpenAlex documentation here: https://docs.openalex.org/api-entities/works/work-object#type

Overall, we see that OpenAlex was most correct on author counts (53.7% of the subset of works), whereas WoS was correct for 33.2% of works in the subset. The listing of consortia or other group identifiers on the byline was a major cause of discrepancies. While both OpenAlex and WoS have other methods for extracting authors from landing pages and full text, publisher-supplied metadata should align with the published work. There is great inconsistency in how these groups are accounted for, how credit is given for authorship, how members are identified, and in the language used to signify how authors work on behalf of or as part of a consortium. In some cases, the complete author list was included, but in other cases, we could not find a list of the group or consortium members. More work needs to be done to document these local practices to promote consistency in author reporting and responsibilities. We also observed that OpenAlex consistently excludes organization names as contributors, even if they are listed in the byline on the landing page and the PDF version of an article. WoS seems to count the group author as a single author consistently.

## 5.2 On the accuracy and internal consistency of metadata and their implication for bibliometrics.

Sometimes there may be an absolute truth against which we can judge the accuracy of a metadata record. For example, a single-author article recorded as having any other number of authors in a database constitutes a true error and was considered as such in our analysis.

However, as we have discussed throughout the different sections of our analysis, a discrepancy between two records representing the same objects in two databases does not necessarily imply the presence of an error in one database or the other. This was particularly evident for publication years and author counts. This offers support to our choice not to consider one database as the gold standard against which the other is to be evaluated. While this may firmly entrench us in a perpetual instance of Segal's law, it is arguably more reasonable to acknowledge that different databases may yield slightly different outcomes that are both valid. Of course, we are not suggesting that a database may not contain more or fewer errors and may not be considered more or less reliable than another. We are merely stating that, in some instances, discrepancies between two databases do not necessarily allow one to draw firm conclusions about the superiority of one over the other in terms of reflecting the reality that they are meant to capture.

On the other hand, if and only if the database providers have clear and transparent metadata policies, then we may be able to assess a database's internal consistency relative to these policies. In our analysis of author counts, for instance, WoS appeared to be very consistent in its treatment of group authors, which were almost always counted as a single author. If that is indeed the intent, then our assessment would be that there is a high level of internal consistency. However, even if that is the case, the possibility remains that this is actually not the intended way of counting authors by WoS, in which case the metadata would be internally consistent but consistently erroneous, which is arguably worse than being wrong only some of the time.

In sum, we would argue that assessing the validity of a metadata element can only be done by comparing it to the full range of possible truths and considering the metadata's consistency with the database's internal policy. Then we can classify any metadata element into one of four categories: 1) internally consistent and valid, 2) internally consistent but invalid, 3) internally inconsistent but valid, and 4) internally inconsistent and invalid. In the absence of clear policies regarding specific metadata elements, it may be challenging to determine what is an issue and what is not. Differences among data sources regarding indexing coverage are, to a degree, largely understood and part of decision-making, and the choice of WoS and Scopus to be selective in the sources they cover contrasts with OpenAlex's focus on comprehensiveness. It is up to the users to understand the tools they use and to make choices that align with the needs of their analyses (Barbour et al., 2025).

It is also important to acknowledge that database providers do not usually create metadata records from scratch but work within an ecosystem of actors and data infrastructures from which their metadata records are sourced. A lack of standardization of practices across providers and a lack of quality control at the journal or publisher level may pose important challenges to aggregators like OpenAlex and WoS. Indeed, Elsevier acknowledged metadata issues as occasional processing errors inherent to large databases (Meester et al., 2016).

Investigations of the coverage and metadata quality of databases matter because these are tools that we use to conduct bibliometric research and assessments. Are validity and consistency sufficient criteria to consider a record reliable in this context? We would argue that it is not. When calculating some of the most common bibliometric indicators, such as the normalized citation score, we want to compare an article with other articles with the same document types (comparable forms of output) that have been published in the same year (comparable time to accumulate citations). Of course, this approach is already imperfect, as an article published on January 1st will be compared with articles published on December 31 of the same year (twelve months apart), but not with articles published on December 31 of the prior year (one day apart). However, these are methodological choices that are made by the bibliometrics community when designing and adopting indicators. Dealing with alternate truths regarding the publication year of an article (e.g., first publication year vs. issue year) is not a choice, but a deviation from the intent behind the use of an indicator. This issue would perhaps be negligible if the database were internally consistent, but our findings suggest that this is not necessarily the case. More than 50,000 discrepancies in publication years occurred because WoS used the online publication year and OpenAlex used the issue year, and nearly 390,000 discrepancies for which it appears to be the reverse.

## 5.3 Limitations and future work

One of the main limitations of our study is its scope. We focused on four metadata elements, but the databases contain many more that we are not considering. Furthermore, we only compared OpenAlex with WoS and not with other databases, and also only considered publications published between 2021 and 2023 in either database. This provides us with a current snapshot of the databases, but no insights into the temporal dynamics of the discrepancies. We were also limited in our investigation of discrepancies in the information easily accessible on the article's landing page and on the PDFs. OpenAlex and WoS draw much of their data from other sources, such as Crossref, which may offer some additional information that could explain some of the discrepancies observed (Eck & Waltman, 2025). Another important limitation is that the OpenAlex database is rapidly evolving, and some of the observed issues may be caused by processes that have since been modified.

This study focused on the nature of the discrepancies and their frequency. We discussed the possible implications that they may have for bibliometric analyses and research evaluation, but more research will be needed to empirically determine if and how the outcomes of bibliometric analyses are impacted at different levels. We also did not investigate disciplinary differences in the discrepancies. Further research should take into account additional metadata elements, examine disciplinary differences, and assess how metadata quality issues in OpenAlex could affect institutional-level metrics and, thus, the results of institutional rankings like the open edition of the Leiden Ranking.

The Paris Conference on Open Research Information and the Barcelona Declaration on Open Research Information emphasize the need for and normalization of open research information. With the tide turning toward open data sources and researchers and institutions embracing OpenAlex and other open data sources and tools, more research will be needed on the quality and coverage of OpenAlex and the other data sources it depends on. This carries implications for OpenAlex, as they look to the research community for feedback on necessary improvements to their metadata, as well as for those conducting research and research evaluation using open sources, who must remain informed of findings related to limitations of the data they use.

# 6. Conflicts of interest

The authors have no conflicts of interest to report.

# 7. Author contributions

Conceptualization (PM), Data curation (PM, GK, MH, PR, SW, RM, RT), Formal analysis (PM), Investigation (PM, MH, GK, RM, PR, RT, SW), Project administration (PM), Visualization (PM), Writing – original draft (PM, MH, PR), Writing – review & editing (PM, MH, GK, RM, PR, RT, SW)

# 8. References

Alonso-Alvarez, P., & Eck, N. J. van. (2024). Coverage and metadata availability of african publications in OpenAlex: A comparative analysis. *arXiv Preprint arXiv:2409.01120*. https://doi.org/10.48550/arXiv.2409.01120

Alperin, J. P., Portenoy, J., Demes, K., Larivière, V., & Haustein, S. (2024). An analysis of the suitability of OpenAlex for bibliometric analyses. *arXiv Preprint arXiv:2404.17663*. https://doi.org/10.48550/arXiv.2404.17663

Álvarez-Bornstein, B., Morillo, F., & Bordons, M. (2017). Funding acknowledgments in the web of science: Completeness and accuracy of collected data. *Scientometrics*, *112*, 1793–1812. https://doi.org/10.1007/s11192-017-2453-4

Barbour, G., Carter, C., Coates, J., Cobey, K. D., Corker, K. S., Gadd, E., Kramer, B., Lawrence, R., Méndez, E., Neylon, C., Pölönen, J., Stern, B., & Waltman, L. (2025). Criteria for bibliographic databases in a well-functioning scholarly communication and research assessment ecosystem. In *Upstream*. https://doi.org/10.54900/d3ck1-skq19

Bordignon, F. (2024). *Is OpenAlex a revolution or a challenge for bibliometrics/bibliometricians?* https://enpc.hal.science/hal-04520837

Céspedes, L., Kozlowski, D., Pradier, C., Sainte-Marie, M. H., Shokida, N. S., Benz, P., Poitras, C., Ninkov, A. B., Ebrahimy, S., Ayeni, P., Filali, S., Li, B., & Larivière, V. (2024). Evaluating the linguistic coverage of OpenAlex: An assessment of metadata accuracy and completeness. *arXiv Preprint arXiv:2409.10633*. https://doi.org/10.48550/arXiv.2409.10633

Clarivate. (n.d.). The history of ISI and the work of eugene garfield. In *The Institute for Scientific Information*. https://clarivate.com/academia-government/the-institute-for-scientific-information/history/

Culbert, J. H., Hobert, A., Jahn, N., Haupka, N., Schmidt, M., Donner, P., & Mayr, P. (2024). Reference coverage analysis of OpenAlex compared to web of science and scopus. *arXiv Preprint arXiv:2401.16359*. https://doi.org/10.48550/arxiv.2401.16359

Donner, P. (2017). Document type assignment accuracy in the journal citation index data of web of science. *Scientometrics*, *113*, 219–236. https://doi.org/10.1007/s11192-017-2483-y

Eck, N. J. van, Visser, M., & Waltman, L. (2024). *Opening up the CWTS Leiden Ranking: Toward a decentralized and open model for data curation*. https://www.leidenmadtrics.nl/articles/opening-up-the-cwts-leiden-ranking-toward-a-decentralized-and-open-model-for-data-curation

Eck, N. J. van, & Waltman, L. (2025). *Crossref as a source of open bibliographic metadata*. https://doi.org/10.31222/osf.io/smxe5_v2

Franceschini, F., Maisano, D., & Mastrogiacomo, L. (2015). Errors in DOI indexing by bibliometric databases. *Scientometrics*, *102*(3), 2181–2186. https://doi.org/10.1007/s11192-014-1503-4

Franceschini, F., Maisano, D., & Mastrogiacomo, L. (2016a). Empirical analysis and classification of database errors in scopus and web of science. *Journal of Informetrics*, *10*(4), 933–953. https://doi.org/10.1016/j.joi.2016.07.003

Franceschini, F., Maisano, D., & Mastrogiacomo, L. (2016b). The museum of errors/horrors in scopus. *Journal of Informetrics*, *10*(1), 174–182. https://doi.org/10.1016/j.joi.2015.11.006

Haunschild, R., & Bornmann, L. (2024). Usage of OpenAlex for creating meaningful global overlay maps of science on the individual and institutional levels. *PLOS ONE*, *19*(12), e0308041. https://doi.org/10.1371/journal.pone.0308041

Haupka, N., Culbert, J. H., Schniedermann, A., Jahn, N., & Mayr, P. (2024). Analysis of the publication and document types in OpenAlex, web of science, scopus, pubmed and semantic scholar. *arXiv Preprint arXiv:2406.15154*. https://doi.org/10.48550/arXiv.2406.15154

Haustein, S., Bowman, T. D., & Costas, R. (2015). When is an article actually published? An analysis of online availability, publication, and indexation dates. *arXiv Preprint arXiv:1505.00796*. https://doi.org/10.48550/arXiv.1505.00796

Liu, W., Hu, G., & Tang, L. (2018). Missing author address information in web of science—an explorative study. *Journal of Informetrics*, *12*(3), 985–997. https://doi.org/10.1016/j.joi.2018.07.008

Liu, W., Huang, M., & Wang, H. (2021). Same journal but different numbers of published records indexed in scopus and web of science core collection: Causes, consequences, and solutions. *Scientometrics*, *126*(5), 4541–4550. https://doi.org/10.1007/s11192-021-03934-x

Maddi, A., & Baudoin, L. (2022). The quality of the web of science data: A longitudinal study on the completeness of authors-addresses links. *Scientometrics*, *127*, 6279–6292. https://doi.org/10.1007/s11192-022-04525-0

Maddi, A., Maisonobe, M., & Boukacem-Zeghmouri, C. (2024). Geographical and disciplinary coverage of open access journals: OpenAlex, scopus and WoS. *arXiv Preprint arXiv:2411.03325*. https://doi.org/10.48550/arXiv.2411.03325

Mazoni, A., & Costas, R. (2024). *Towards the democratisation of open research information for scientometrics and science policy: the Campinas experience*. https://www.leidenmadtrics.nl/articles/towards-the-democratisation-of-open-research-information-for-scientometrics-and-science-policy-the-campinas-experience

Meester, W. J. N., Colledge, L., & Dyas, E. E. (2016). A response to "the museum of errors/horrors in scopus" by franceschini et al. *Journal of Informetrics*, *10*(2), 569–570. https://doi.org/10.1016/j.joi.2016.04.011

Narin, F. (1976). *Evaluative bibliometrics: The use of publication and citation analysis in the evaluation of scientific activity*. Computer Horizons Washington, D. C.

Ortega, J. L., & Delgado-Quirós, L. (2024). The indexation of retracted literature in seven principal scholarly databases: A coverage comparison of dimensions, OpenAlex, PubMed, scilit, scopus, the lens and web of science. *Scientometrics*, *129*(7), 3769–3785. https://doi.org/10.1007/s11192-024-05034-y

Priem, J., Piwowar, H., & Orr, R. (2022). OpenAlex: A fully-open index of scholarly works, authors, venues, institutions, and concepts. *arXiv Preprint arXiv:2205.01833*. https://doi.org/10.48550/arxiv.2205.01833

Schares, E. (2024). Comparing funder metadata in OpenAlex and dimensions. *Iowa State University*. https://doi.org/10.31274/b8136f97.ccc3dae4

Shi, J., Nason, M., Tullney, M., & Alperin, J. (2025). Identifying metadata quality issues across cultures. *College & Research Libraries*, *86*(1). https://doi.org/10.5860/crl.86.1.101

Simard, M.-A., Basson, I., Hare, M., Lariviere, V., & Mongeon, P. (2024). The open access coverage of OpenAlex, scopus and web of science. *arXiv Preprint arXiv:2404.01985*. https://doi.org/10.48550/arXiv.2404.01985

Simard, M.-A., Basson, I., Hare, M., Larivière, V., & Mongeon, P. (2025). Examining the geographic and linguistic coverage of gold and diamond open access journals in OpenAlex, scopus and web of science. *Quantitative Science Studies*, 1–29. https://doi.org/10.1162/QSS.a.1

Singh, P., Singh, V. K., & Kanaujia, A. (2024). Exploring the publication metadata fields in web of science, scopus and dimensions: Possibilities and ease of doing scientometric analysis. *Journal of Scientometric Research*, *13*(3), 715–731. https://doi.org/10.5530/jscires.20041144

Visser, M., Eck, N. J. van, & Waltman, L. (2021). Large-scale comparison of bibliographic data sources: Scopus, web of science, dimensions, crossref, and microsoft academic. *Quantitative Science Studies*, *2*(1), 20–41. https://doi.org/10.1162/qss_a_00112

Zhang, L., Cao, Z., Shang, Y., Sivertsen, G., & Huang, Y. (2024). Missing institutions in OpenAlex: Possible reasons, implications, and solutions. *Scientometrics*. https://doi.org/10.1007/s11192-023-04923-y